\documentclass[aps,reprint,superscriptaddress,longbibliography,floatfix]{revtex4-1}

\usepackage[utf8]{inputenc}
\usepackage{hyperref}
\usepackage{graphicx}
\usepackage{amsmath}
\usepackage{bm}
\usepackage{amssymb}
\usepackage{soul}

\usepackage{color}

\begin{document}

\title{First-order magnetic phase-transition of mobile electrons in monolayer MoS$_{2}$}
\author{Jonas G. Roch}
\email{jonasgael.roch@unibas.ch}
\affiliation{Department of Physics, University of Basel, Klingelbergstrasse 82, CH-4056 Basel, Switzerland}

\author{Dmitry Miserev}
\affiliation{Department of Physics, University of Basel, Klingelbergstrasse 82, CH-4056 Basel, Switzerland}

\author{Guillaume Froehlicher}
\affiliation{Department of Physics, University of Basel, Klingelbergstrasse 82, CH-4056 Basel, Switzerland}

\author{Nadine Leisgang}
\affiliation{Department of Physics, University of Basel, Klingelbergstrasse 82, CH-4056 Basel, Switzerland}

\author{Lukas Sponfeldner}
\affiliation{Department of Physics, University of Basel, Klingelbergstrasse 82, CH-4056 Basel, Switzerland}

\author{Kenji Watanabe}
\affiliation{National Institute for Material Science, 1-1 Namiki, Tsukuba, 305-0044, Japan}

\author{Takashi Taniguchi}
\affiliation{National Institute for Material Science, 1-1 Namiki, Tsukuba, 305-0044, Japan}

\author{Jelena Klinovaja}
\affiliation{Department of Physics, University of Basel, Klingelbergstrasse 82, CH-4056 Basel, Switzerland}

\author{Daniel Loss}
\affiliation{Department of Physics, University of Basel, Klingelbergstrasse 82, CH-4056 Basel, Switzerland}

\author{Richard J. Warburton}
\affiliation{Department of Physics, University of Basel, Klingelbergstrasse 82, CH-4056 Basel, Switzerland}

\begin{abstract}
Evidence is presented for a first-order magnetic phase transition in a gated two-dimensional semiconductor, monolayer-MoS$_{2}$. The phase boundary separates a spin-polarised (ferromagnetic) phase at low electron density and a paramagnetic phase at high electron density. Abrupt changes in the optical response signal an abrupt change in the magnetism. The magnetic order is thereby controlled via the voltage applied to the gate electrode of the device. Accompanying the change in magnetism is a large change in the electron effective mass.
\end{abstract}

\maketitle

Mobile electrons in a semiconductor can lower their energy by aligning their spins, a consequence of the Pauli principle. A ferromagnetic phase in which all electron spins point in the same direction was proposed originally by Bloch \cite{Bloch1929}. Experimental verification of this prediction on two-dimensional (2D) electron gases in conventional semiconductors was elusive \cite{Zhu2003,Vakili2004} on account of disorder \cite{Finkelstein1995,Yusa2000}. Recently however, ferromagnetic ordering of 2D electrons was reported in monolayer MoS$_{2}$ \cite{Roch2019} and twisted-bilayer graphene \cite{Sharpe2019}. A phase transition can be expected between a ferromagnetic state at low density and a paramagnetic state at high density. The behaviour on crossing the phase boundary is crucial: does the magnetisation turn on abruptly or does the magnetisation increase gradually? 

In three-dimensional (3D) ferromagnets such as iron, the magnetisation grows gradually from zero on crossing the phase boundary. This is classified as a ``second-order" phase transition. An abrupt change in the magnetisation is more striking and potentially much more useful in spintronics: a small change in a control parameter from one side of the phase boundary to the other results in a massive change in the magnetisation. An abrupt change is classified as a ``first-order" phase transition. In metallic systems, the control parameter is typically the temperature or pressure, neither convenient for fast and efficient switching from one phase to the other. This restriction is lifted in 2D semiconductors for which the electron density can be controlled over a wide range simply via a voltage applied to a gate electrode. However, magnetism of mobile electrons in 2D is different to that in 3D. On the one hand, mean-field theories, such as those of Bloch \cite{Bloch1929} and Stoner \cite{Stoner1938}, are provably inadequate in 2D \cite{Mermin1966,Loss2011}. On the other hand, corrections to Fermi liquid theory are predicted to be much more pronounced in 2D compared to 3D \cite{Belitz1999,Chubukov2004,Maslov2009,Brando2016}. Experimentally, magnetic phase transitions of mobile electrons in 2D semiconductors are unexplored. 

\begin{figure}[t]
\centering
\includegraphics[width=80mm]{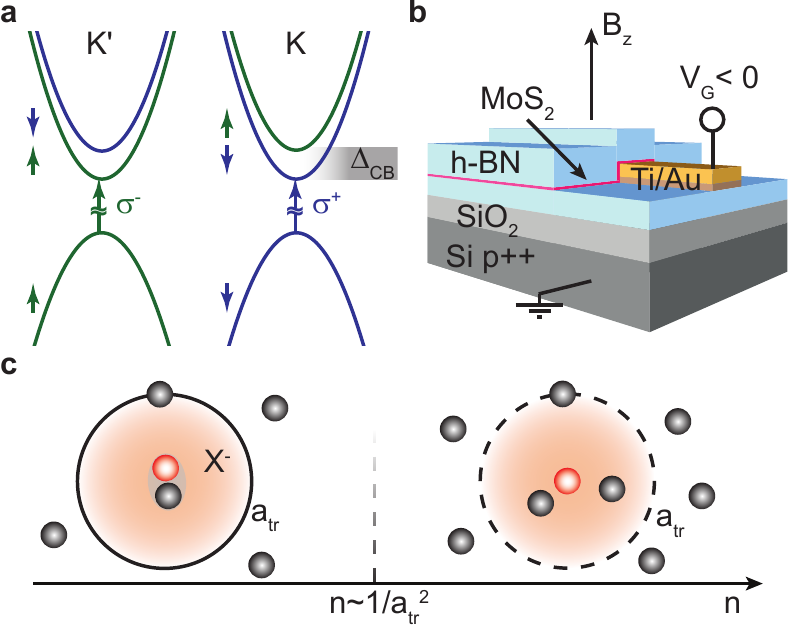}
\caption{(a) Band structure and allowed optical transitions of monolayer MoS$_2$. (b) A van der Waals heterostructure consisting of monolayer MoS$_2$  embedded in h-BN. The doped silicon substrate is earthed and a voltage is applied to a Ti/Au layer in contact with the MoS$_{2}$. A magnetic field is applied perpendicular to the layers. (c) Schematic representation of a trion in a two-dimensional Fermi sea. The black circles denote conduction-band electrons; the red circle a valence-band hole. The trion (X$^{-}$, Bohr radius $a_{\rm tr}$) is bound (unbound) at low (high) electron density.}
\label{fig1}
\end{figure}

We focus on monolayer MoS$_{2}$, a 2D semiconductor in the transition-metal dichalcogenide (TMDC) family. We present evidence for a first-order phase transition between a paramagnetic phase at high electron-density and a ferromagnetic phase at low electron-density. The magnetism is thereby controlled electrically simply via the voltage applied to a gate electrode. Accompanying this abrupt change in magnetisation is an abrupt change in the electron effective mass, from a relatively small value in the paramagnetic phase to a large value in the ferromagnetic phase. 

\begin{figure*}[t]
\centering
\includegraphics[width=0.8\textwidth]{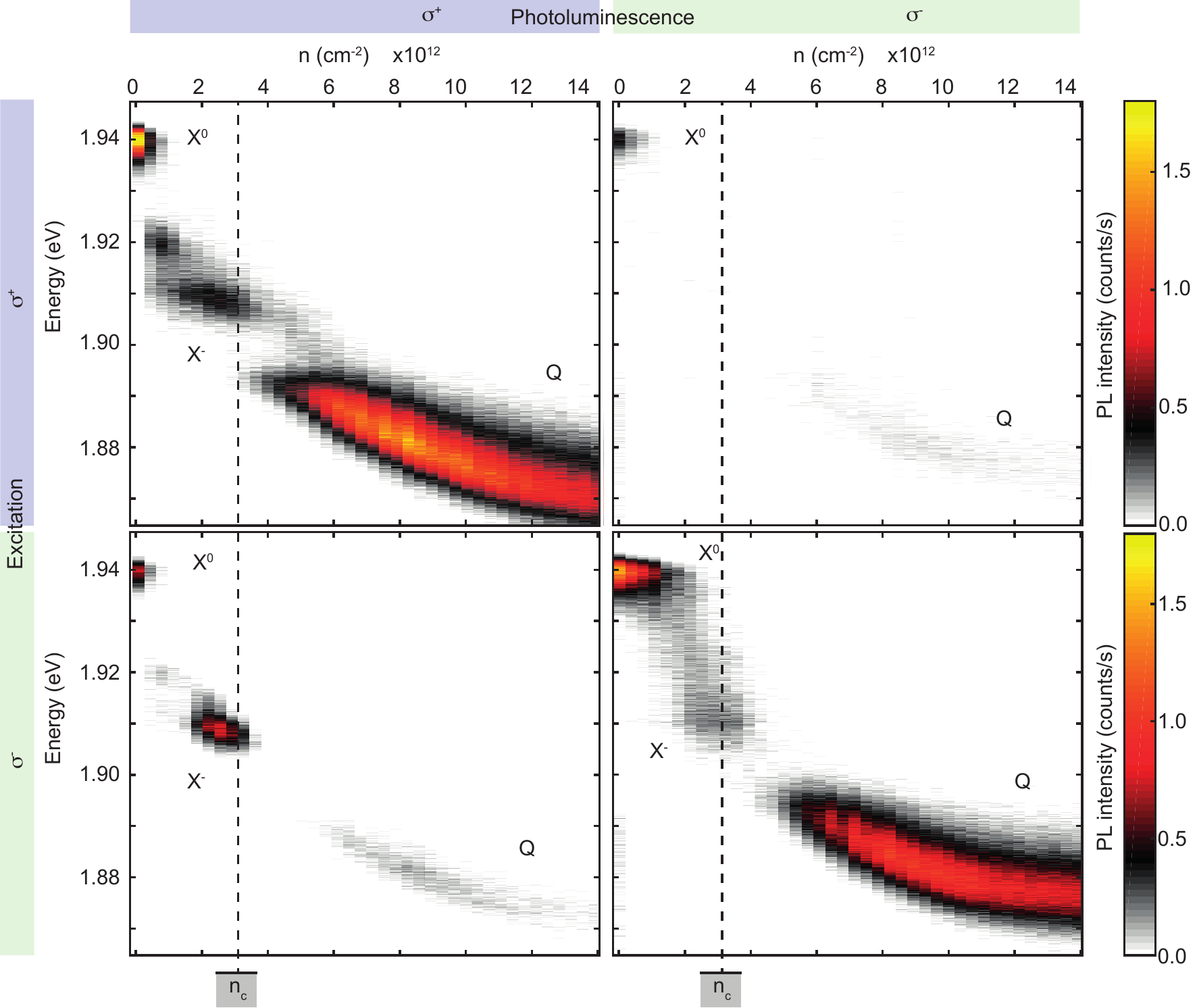}
\caption{Photoluminescence of gated monolayer-MoS$_{2}$. The excitation is either $\sigma^{+}$ or $\sigma^{-}$; the detection either $\sigma^{+}$ or $\sigma^{-}$. Data recorded at 1.6 K and $B_{z}=9.00$ T $n_{\rm c}$ denotes the critical density. At $n=n_{\rm c}$ there is a jump in the PL energy: the nature of the emission changes abruptly from a trion (X$^{-}$) to a Mahan exciton (Q). For $n>n_{\rm c}$, the PL process is largely polarisation conserving. Conversely, for $n<n_{\rm c}$, the PL is highly polarisation nonconserving: $\sigma^{-}$ excitation leads to strong $\sigma^{+}$ PL.}
\label{fig2}
\end{figure*}

The conduction band structure of MoS$_{2}$ exhibits minima at the edges of the Brillouin zone, at the K and K$^{\prime}$ points \cite{Mak2010,Xiao2012,Kormanyos2014}. The spin-$\uparrow$ and spin-$\downarrow$ states are split by a small spin-orbit interaction, $\Delta_{\rm CB}$ (Fig.\ 1a). Calculations predict $\Delta_{\rm CB}=3$ meV \cite{Kormanyos2014}; experiments suggest $\Delta_{\rm CB}=0.8$ meV \cite{Marinov2017}. The bare electron mass is relatively large, $0.44 m_{\rm o}$ according to theory \cite{Kormanyos2014}; the dielectric constant relatively small, such that the Bohr radius is just $\sim 0.5$ nm, only slightly larger than the lattice constant. Quantum effects in gated MoS$_{2}$ have recently been reported: conductance quantisation \cite{Marinov2017}, Shubnikov-de Haas oscillations \cite{Pisoni2018} and a spontaneous spin-polarisation \cite{Roch2019}.

Here, a monolayer of MoS$_{2}$ is embedded between two h-BN layers. Electrons are injected into the MoS$_{2}$ by applying a voltage to a metallic contact (Fig.\ 1b). The ground state of the mobile electrons is probed optically with photoluminescence (PL) and absorption spectroscopy. In the PL experiment, a laser with photon energy 1.959 eV is just slightly blue-detuned with respect to the optical transition. Right-handed circularly-polarised light ($\sigma^{+}$) injects a spin-$\downarrow$ hole at the K-point; left-handed circularly-polarised light ($\sigma^{-}$) injects a spin-$\uparrow$ hole at the K$^{\prime}$-point (Fig.\ 1a). This spin-valley selectivity is a crucial aspect: the optical probe represents a spin-sensitive probe of the electronic ground-state. Furthermore, an optical experiment represents a local measurement: with confocal microscopy, information is gleaned from a few-hundred nanometre diameter region on the sample. On this length scale, inhomogeneous broadening in the optical response is small \cite{Cadiz2017,Ajayi2017}. Specifically, the experiment consists of cooling to 1.6 K, applying a magnetic field of $B_{z}=9.0$ T (to create a stable spin polarisation \cite{Roch2019}) and measuring polarisation-resolved PL as a function of the voltage applied to the electrode, equivalently the electron density ($n$).

PL in the presence of a Fermi sea of electrons has been explored in a number of 2D semiconductor systems (GaAs \cite{Finkelstein1995,Yusa2000}, CdTe \cite{Huard2000}, MoSe$_{2}$ \cite{Sidler2016,Back2017}) and there is a model to describe it \cite{Uenoyama1990,Hawrylak1991}. At very low $n$, PL arises from the recombination of tightly bound electron-hole pairs, excitons (X$^{0}$). As $n$ increases, X$^{0}$ weakens and is replaced with a red-shifted peak, the trion (X$^{-}$). The trion consists of two electrons in a spin-singlet state and a hole, or, in more accurate language, the bound exciton-Fermi sea polaron \cite{Suris2001,Efimkin2017}. At higher $n$, the trions become unbound. In a simple picture, this occurs once an electron from the Fermi sea resides ``inside" the trion wave-function (Fig.\ 1c). In the PL spectrum, the X$^{-}$ peak evolves gradually into a broad peak, typically red-shifted with respect to X$^{-}$, the Mahan exciton \cite{Mahan1967}. 
A key property of the Mahan exction is the red-shift of PL with respect to absorption, $E_{\rm PL}-E_{\rm A}<0$ \cite{Uenoyama1990,Hawrylak1991}. In the single-particle limit for equal electron and hole masses (as for TMDCs), this shift is twice the Fermi energy. 

\begin{figure}[t]
\centering
\includegraphics[width=80mm]{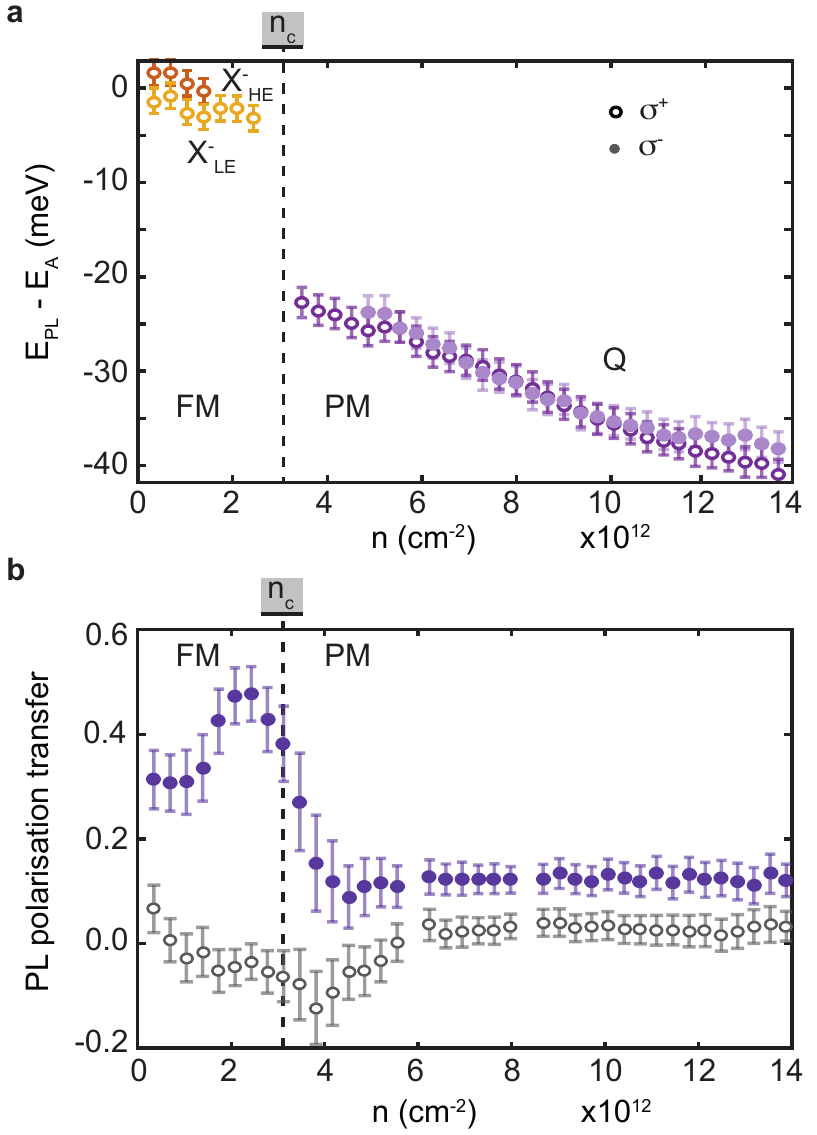}
\caption{(a) The energetic separation between the PL and the absorption, $E_{\rm PL}-E_{\rm A}$ (measured either with $\sigma^{+}$ or $\sigma^{-}$ polarisation), versus electron density at 1.6 K and $B_{z}=9.00$ T. There are two trions, X$^{-}_{\rm LE}$ and X$^{-}_{\rm HE}$ (red and orange points, respectively) \cite{Roch2019}. The Q-peak is plotted for both polarisations ($\sigma^{+}$ and $\sigma^{-}$, open-purple and filled-purple points, respectively). (b) The PL polarisation transfer [$a_{2}/(a_{1}+a_{2}+a_{3})$ (purple) and $a_{3}/(a_{1}+a_{2}+a_{3})$ (white), see eq.\ \ref{PLmatrix}], versus electron density. These plots determine $n_{\rm c}=3.0 \times 10^{12}$ cm$^{-2}$.}
\label{fig3}
\end{figure}

We turn to the experiment on gated MoS$_{2}$. The absorption spectra at large magnetic field (two trions in $\sigma^{+}$-polarisation but no trions in $\sigma^{-}$-polarisation) show that the spins are polarised at low-to-intermediate densities \cite{Roch2019}. We focus here on PL. At $n \simeq 0$, the PL spectrum consists of a single sharp line corresponding to emission from excitons (X$^{0}$ in Fig.\ 2) \cite{Chernikov2014}. On increasing $n$, X$^{0}$ weakens and is replaced with a red-shifted line (X$^{-}$ in Fig.\ 2), the trion \cite{Sidler2016,Back2017,Roch2019}. At higher densities, X$^{-}$ disappears and a broad peak (labelled Q in Fig.\ 2) appears to the red. The Q-peak exhibits the large PL-absorption splitting characteristic of the Mahan exciton. These features follow the standard behaviour of PL in the presence of a Fermi sea. There is however a radically different feature. Strikingly, the smooth transition between trion (X$^{-}$) and Mahan excxiton (Q) is missing: instead, there is an abrupt change from one to the other resulting in a ``gap" in the PL spectrum (Fig.\ 2). This signals an abrupt change in the Fermi sea: it is the first evidence for a first-order phase transition. The second evidence comes from the polarisation of the PL. We excite and detect in all four combinations of circular polarisation, presenting the results as a matrix (Fig.\ 2). For the Q-peak, the response is overwhelmingly ``diagonal", i.e.\ polarisation-preserving: $\sigma^{+}$-polarised excitation results in $\sigma^{+}$-polarised PL; $\sigma^{-}$-polarised excitation results in $\sigma^{-}$-polarised PL. Conversely, for X$^{-}$, there is a large ``non-diagonal" response ($\sigma^{-}$-polarised excitation results in $\sigma^{+}$-polarised PL yet $\sigma^{+}$-polarised excitation results in a very weak PL signal), clear evidence for symmetry breaking. The switch from one symmetry class to the other is also abrupt as the density changes.

To make a quantitative analysis, we plot $E_{\rm PL}-E_{\rm A}$ as a function of $n$ (Fig.\ 3a). There is an abrupt change at $3.0 \times 10^{12}$ cm$^{-2}$ signifying an abrupt change in the nature of the PL process. Also, we analyse the polarisation dependence via:
\begin{equation}
\label{PLmatrix}
\begin{pmatrix}
P_{+} \\
P_{-} \\
\end{pmatrix}=
\begin{pmatrix}
a_{1} + h & a_{2} \\
a_{3} & a_{1} -h \\
\end{pmatrix}
\begin{pmatrix}
L_{+} \\
L_{-} \\
\end{pmatrix}
\end{equation}
where $L_{+}$ is the $\sigma^{+}$-polarised laser intensity, $L_{-}$ the $\sigma^{-}$-polarised laser intensity, $P_{+}$ the $\sigma^{+}$-polarised PL signal, and $P_{-}$ the $\sigma^{-}$-polarised PL signal. $P_{+}$ and $P_{-}$ are integrated over the spectral window (Fig.\ 2). The term $a_{1}$ describes the polarisation-preserving response. Terms $a_{2}$ and $a_{3}$ describe a transfer of the polarisation, from $\sigma^{-}$ at the input to $\sigma^{+}$ at the output ($a_{2}$); from $\sigma^{+}$ at the input to $\sigma^{-}$ at the output ($a_{3}$). The terms $a_{1}$, $a_{2}$ and $a_{3}$ are the important ones here. There is one further term. In the experiment, a hole is injected either at the K-point or at the K$^{\prime}$-point by choosing the polarisation (Fig.\ 1a). This works well but imperfectly. In an applied magnetic field, the hole can lower its energy by relaxing from the K-point to the K$^{\prime}$-point. This process is described by term $h$. We define the ``PL polarisation transfer" as $a_{2}/(a_{1}+a_{2}+a_{3})$, i.e.\ the fraction of the total PL emitted in the polarisation-nonconserving channel. As a function of $n$, the PL polarisation transfer reaches values as high as 50\% at intermediate density, decreasing rapidly between $n=2.5 \times 10^{12}$ and $n=3.5 \times 10^{12}$ cm$^{-2}$ (Fig.\ 3b). The opposite polarisation-nonconserving process, $a_{3}/(a_{1}+a_{2}+a_{3})$, is small at all $n$ (Fig.\ 3b). Both $E_{\rm PL}-E_{\rm A}$ and the PL polarisation transfer change abruptly at the same electron density. We identify a critical density of $n_{\rm c}=3.0 \times 10^{12}$ cm$^{-2}$ (Fig.\ 3).

\begin{figure}[t]
\centering
\includegraphics[width=80mm]{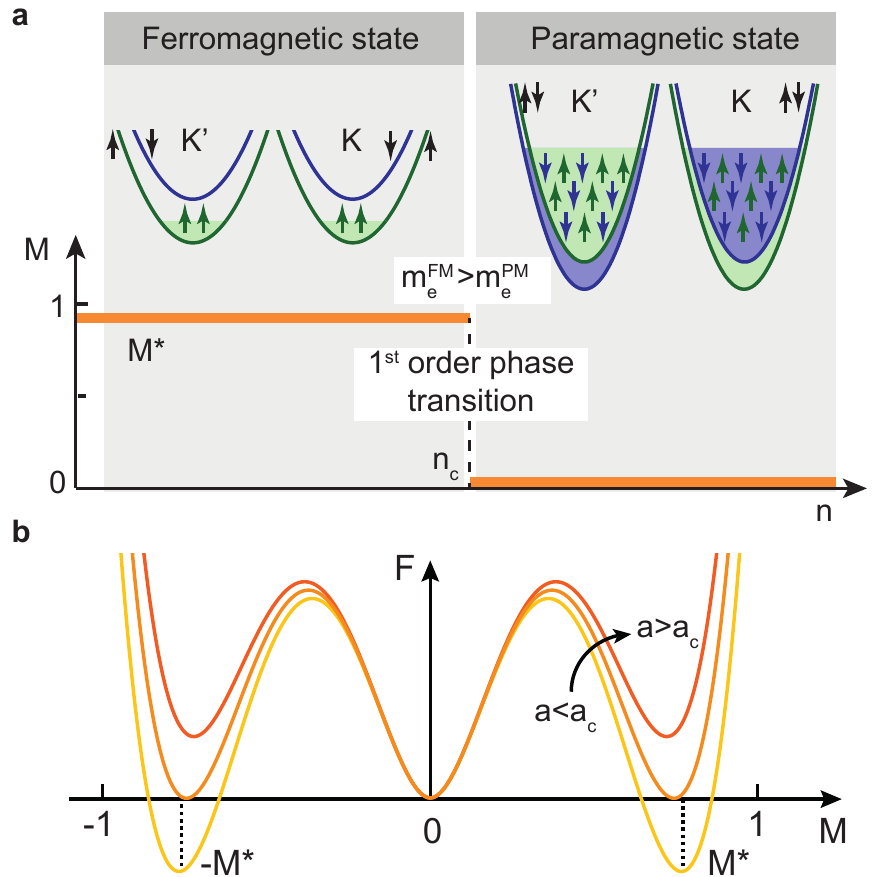} 
\caption{(a) The occupied bands in the ferromagnetic phase (left) and in the paramagnetic phase (right) along with the change in magnetisation at $n=n_{\rm c}$. In the ferromagnetic phase, occupation of two bands with the same spin is favoured, a consequence of exchange. In the paramagnetic phase, all four bands are occupied. (b) The free energy $F$ as a function of magnetisation fraction $M$ for $F=aM^{2}+bM^{4}+c|M|^{3}$. The terms $aM^{2}$ and $bM^{4}$ are the terms from Ginzburg-Landau theory; the nonanalytic term $c |M|^{3}$ arises from corrections to Landau's theory of the Fermi liquid. Plotted is $F$ for constant $b$ and $c$ with $c<0$ as a function of $a$. For $a<c^{2}/4b$, the global minimum of $F$ lies at $M=0$ corresponding to a paramagnetic phase. For $a>c^{2}/4b$, the global minimum of $F$ lies at large $M$ corresponding to a ferromagnetic phase. The phase transition at $a=c^{2}/4b$ is first-order.}
\label{fig4}
\end{figure}

The experiment shows that $E_{\rm PL}-E_{\rm A}$ and the PL polarisation transfer are excellent probes of the magnetic order in this system. The PL polarisation transfer gives an immediate measure of the symmetry breaking in the system: it is large in the ferromagnetic phase, small in the paramagnetic phase. Unfortunately, it is a very complex task to describe the entire density dependence of the PL process theoretically, and a model specific to gated MoS$_{2}$ does not yet exist. Nevertheless, we use these two features to explore the dependence of the phase transition on magnetic field and temperature ($T$). At low-$T$ and $B_{z}=0$, the abrupt change in $E_{\rm PL}-E_{\rm A}$ is still clearly visible. This shows that even at $B_{z}=0$, there is still a phase transition at $n=n_{\rm c}$ between states with different magnetic order. However, at $B_{z}=0$ the two PL polarisation transfer terms are the same (see Supplementary Section III). (For $n<n_{\rm c}$ and $B_{z}=+9.00$ T, $a_{2} \gg a_{3}$; for $n<n_{\rm c}$ and $B_{z}=0$, $a_{2}=a_{3}$.) This tells us that, averaged over time and space, the magnetisation is zero. To be consistent with the PL, the magnetisation must be locally ferromagnetic: for $n<n_{\rm c}$, there are fluctuating ``puddles" of spin-$\uparrow$ electrons and ``puddles" of spin-$\downarrow$ electrons with zero average magnetisation; for $n>n_{\rm c}$ the magnetisation disappears completely. The behaviour for $n<n_{\rm c}$ is consistent with the Mermin-Wagner theorem \cite{Mermin1966}. For $n<n_{\rm c}$, application of a magnetic field stabilises one spin orientation. In a large $B_{z}$ but at $T=30$ K, the abrupt change in $E_{\rm PL}-E_{\rm A}$ is also clearly visible but the PL polarisation transfer reduces with respect to 1.6 K (see Supplementary Section III). This suggests that the magnetic phase transition survives at 30 K, but that in the ferromagnetic phase, thermal fluctuations between spin-$\uparrow$ and spin-$\downarrow$ states become important. It is an open question if the magnetic phase transition survives to even higher temperatures.

Our results are in agreement with recent theory \cite{Miserev2019} which predicted spin-ordering (and not ordering in the valley index or combined spin-valley index \cite{Donck2018,Braz2018}) and a first-order phase transition between ferromagnetic and paramagnetic phases. Both predictions in the theory depend on corrections to Fermi liquid theory which arise via infrared electron-hole excitations at the Fermi energy \cite{Belitz1999,Chubukov2004,Maslov2009}. This theory allows us to present a model which is fully consistent with the experimental results (Fig.\ 4). For $n<n_{\rm c}$, the two bands with spin-$\uparrow$ are occupied resulting in a spin-polarisation \cite{Roch2019}; for $n>n_{\rm c}$, all four bands are occupied and the spin polarisation disappears (Fig.\ 4a). The free energy $F$ depends on the magnetisation fraction $M$ according to:
\begin{equation}
F=a M^{2}+b M^{4} +c |M|^{3}.
\end{equation}
The first two terms represent the Ginzburg-Landau model which, alone, lead to a second-order phase transition. The third term is a non-analytic correction. Crucially, theory predicts a negative $c$ for MoS$_{2}$ \cite{Miserev2019}. Plots of $F$ as a function of density (i.e.\ versus $a$) show how a negative $c$ leads to a first-order transition from a paramagnetic phase to a ferromagnetic phase (Fig.\ 4b). We stress that the first-order nature of the phase transition depends on the non-analytic correction: without it, the transition would be second-order.

In the ferromagnetic (FM) phase, we observe PL from the trion: the trion is bound; in the paramagnetic (PM) phase we do not observe PL from the trion: the trion is unbound. In other words, there is an abrupt change in the nature of the PL at $n_{\rm c}$, from trion to Mahan exciton. This implies a profound change of the electron mass on passing from the FM- to the PM-phase. We analyse the simple concept (Fig.\ 1c). The Fermi sea reduces the trion binding energy $E_{\rm T}$ because all states with $k<k_{\rm F}$ are occupied and become unavailable in constructing the trion wave-function: $E_{\rm T}=E_{\rm T}^{0}-\hbar^2 k_{\rm F}^2/2\mu$. ($k_{\rm F}$ is the Fermi wave-vector, $E_{\rm T}^{0}$ the trion binding energy in the limit $n \rightarrow 0$, and $\mu$ the reduced mass of an exciton and an electron. This result comes from the Suris model \cite{Suris2001} with the approximation $E_{\rm T}^{0} \gg \hbar^{2} k_{\rm F}^2/2\mu$.) In the FM [PM] phase, two [four] bands are occupied such that $E_{\rm T}=0$ at $n_{\rm FM}=\mu E_{\rm T}^{0}/\pi \hbar^{2}$ [$n_{\rm PM}=2\mu E_{\rm T}^{0}/\pi \hbar^{2}$]. The experiment tells us that $n_{\rm FM}$ must be larger than $n_{\rm c}$, and that $n_{\rm PM}$ must be smaller than $n_{\rm c}$, i.e.\ $n_{\rm FM}>n_{\rm PM}$. This conundrum can only be resolved by a change in the reduced mass, $\mu_{\rm FM}>2\mu_{\rm PM}$, i.e.\ a significant decrease in the electron mass on going from the FM to the PM phase. Taking $E_{\rm T}^{0}=17$ meV \cite{Roch2019} and a hole mass of $m_{\rm h}=0.54 m_{\rm o}$ \cite{Kormanyos2015}, we find $m_{\rm e}^{\rm FM}>0.65 m_{\rm o}$ and $m_{\rm e}^{\rm PM}<0.40 m_{\rm o}$. The mass in the ferromagnetic phase ($m_{\rm e}^{\rm FM}$) is consistent with the value deduced from Shubnikov-de Haas oscillations at low density ($0.8 m_{\rm o}$) \cite{Pisoni2018}; the mass in the paramagnetic phase ($m_{\rm e}^{\rm PM}$) is consistent with the calculated bare electron mass ($0.44 m_{\rm o}$) \cite{Kormanyos2014}. This analysis is consistent with general expectations that the mass is enhanced above its bare value in an interacting phase.

As an outlook, we comment that, first, our work establishes MoS$_{2}$ as a platform for studying and exploiting interaction-driven physics. The combination of an ultra-small Bohr radius, spin and valley degrees of freedom, and a small spin-orbit interaction creates a rich test bed; the direct band-gap allows these states to be imprinted onto the optical properties. Secondly, the first-order phase transition between ferromagnetic and paramagnetic phases enables the spin ordering to be controlled simply via a small change to a gate voltage opening a route to fast and efficient electrical switching. The next step is to stabilise the ferromagnetic phase without a large external field, for instance via coupling to an insulating 2D ferromagnet in a van der Waals heterostructure. 

We thank Mansour Shayegan for fruitful discussions. The work was supported by SNF (Project No. 200020\_156637), the Georg H.\ Endress foundation, QCQT, SNI (Swiss Nanoscience Institute) and NCCR QSIT. K.W.\ and T.T.\ acknowledge support from the Elemental Strategy Initiative conducted by the MEXT, Japan and the CREST(JPMJCR15F3), JST.

%


\newpage
\begin{figure*}[t]
\centering
\includegraphics[width=150mm]{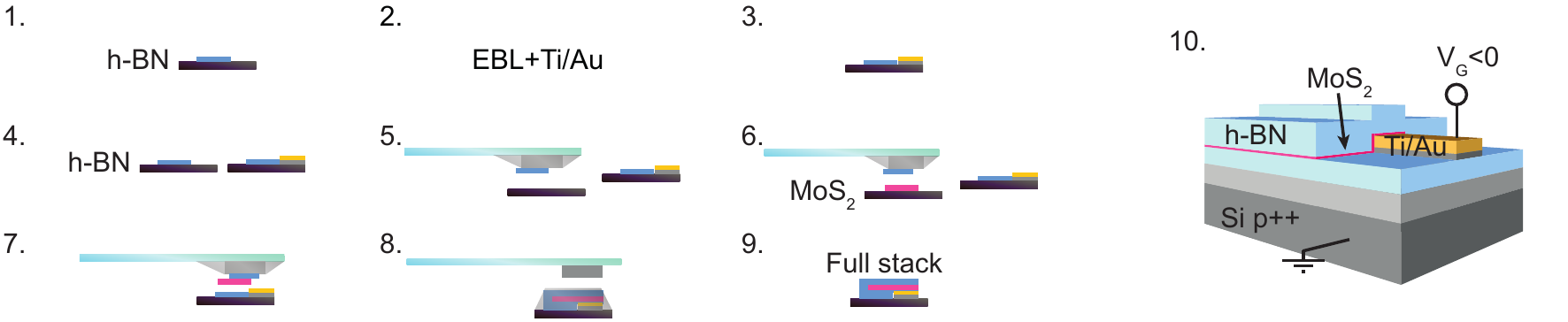}
\caption{Device fabrication. (1) h-BN is exfoliated onto a heavily p-doped Si/SiO$_2$ substrate. (2+3) Using electron-beam lithography (EBL), electric contacts are defined on the h-BN flake of interest and the metallic contacts (Ti/Au) are subsequently evaporated. (4+5) More h-BN is exfoliated on another chip and a flake is picked up. (6+7) A MoS$_2$ monolayer is picked up using the h-BN already on the stamp. (7) The top h-BN and MoS$_2$ flakes are placed on the bottom h-BN with the metallic contacts. (8) The polycarbonate (PC) film is melted; the van der Waals heterostructure stays on the chip along with the metallic contacts. (9+10) The PC is dissolved in chloroform and the sample fabrication is complete.}
\label{stackingProcess}
\end{figure*}
\section*{Supplementary information}

\begin{figure}[h]
\centering
\includegraphics[width=83mm]{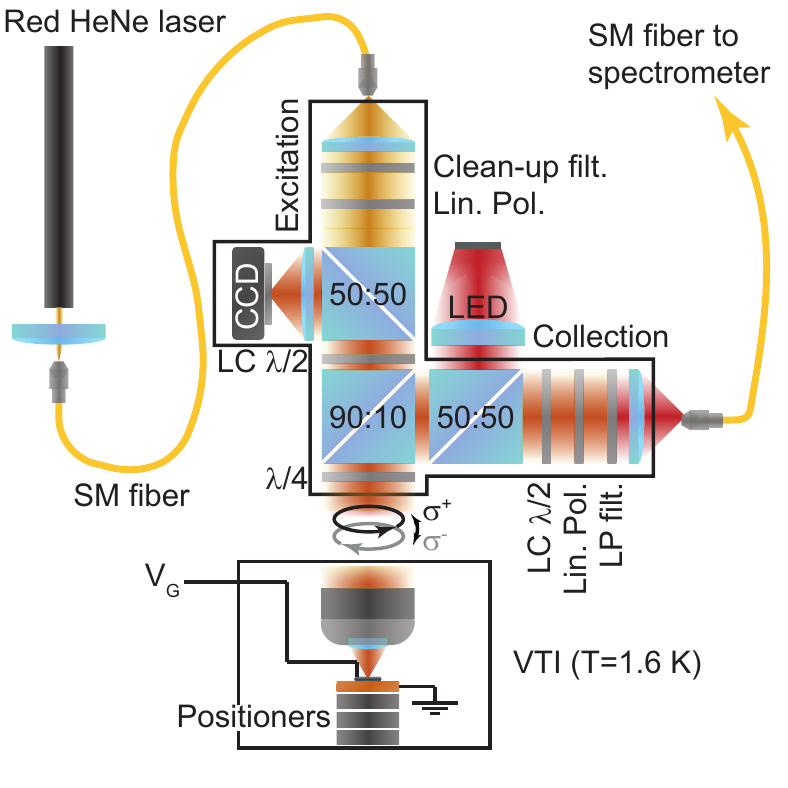}
\caption{Schematic of the cryogenic confocal-microscope. SM stands for single-mode fibre; VTI for variable-temperature insert; LC for liquid crystal; LP for low-pass filter.}
\label{upl}
\end{figure}
\section{Sample fabrication}
Van der Waals heterostructures were fabricated using a dry-transfer technique~\cite{Zomer2014}. A thin layer of polycarbonate (PC) placed on top of a polydmethylsiloxane (PDMS) stamp is used to pick up flakes exfoliated on SiO$_2$(300~nm)/Si substrates. Bulk crystals were used to carry out exfoliation (natural MoS$_2$ from SPI Supplies, synthetic h-BN~\cite{Taniguchi2007}, and natural graphite from NGS Naturgraphit). Bottom metal Au(45~nm)/Ti(5~nm) contacts were patterned by electron-beam lithography (EBL) in the middle of the stacking process, as depicted in Fig.~\ref{stackingProcess}, following Ref.~\cite{Pisoni2018}.

The electron density in the sample is deduced from the geometrical capacitance $C$ given by the capacitor formed between the monolayer MoS$_2$ and the p-doped silicon substrate,
\begin{equation}
C=\dfrac{1}{\dfrac{d_{\rm BN}}{\epsilon_{\rm BN}}+\dfrac{d_{\rm SiO_2}}{\epsilon_{\rm SiO_2}}}~,
\end{equation}
where  $\epsilon_{\rm BN}=3.76$ and $\epsilon_{\rm SiO_2}=3.9$ are the dielectric constant of h-BN and SiO$_2$, respectively, and $d_{\rm SiO_2}=300$~nm and $d_{\rm BN}$ are the oxide and h-BN thicknesses. Using these parameters, we obtain the geometrical capacitance $C=11.1\pm 0.5$~nFcm$^{-2}$ where a 5\% uncertainty in the layer thicknesses is taken into account.
\section{Experimental setup}
\subsection{Photoluminescence measurement}

Photoluminescence (PL) was measured using a home-built confocal microscope, as shown in Fig.~\ref{upl}. A variable-temperature insert (VTI) within a helium bath-cryostat allows the temperature to be controlled. With our setup, the sample can reach a temperature as low as $T=1.6$~K. A red helium-neon laser ($\lambda=633$~nm) is used for the laser excitation. The laser light is coupled into a single-mode optical fibre. The light is out-coupled from the fibre and enters the excitation arm of the microscope (Fig.~\ref{upl}). Any background light is suppressed using a narrow-band band-pass (``clean-up") filter. A linear polariser sets the polarisation axis of the excitation beam. A computer-controlled liquid-crystal variable retarder is subsequently used in combination with a $\lambda/4$ wave-plate such that the polarisation of the laser can be set to the left- or right-handed circular polarisation state. The light is then focussed to a diffraction-limited spot on the sample's surface using a single aspherical lens (numerical aperture 0.7). The PL excitation intensity is kept low, such that no more than 500~nW reaches the sample. The focus and position of the spot on the sample can be controlled {\em in situ} using piezoelectric nano-positioners.  

The PL signal is collected through the same aspherical lens and is sent to the collection arm (Fig.~\ref{upl}) by a beam-splitter. The polarisation state entering the detection fibre using a $\lambda/4$ wave-plate, a computer-controlled liquid-crystal variable-retarder and a linear polariser (Fig.~\ref{upl}). Any residual laser light is attenuated using a long-pass filter. The PL signal is coupled into a single-mode fibre connected to a spectrometer. The PL is then dispersed by a 1500~g/mm grating and focussed onto a nitrogen-cooled CCD camera.

\subsection{Gate sweeps}

\begin{figure*}[t]
\centering
\includegraphics[width=150mm]{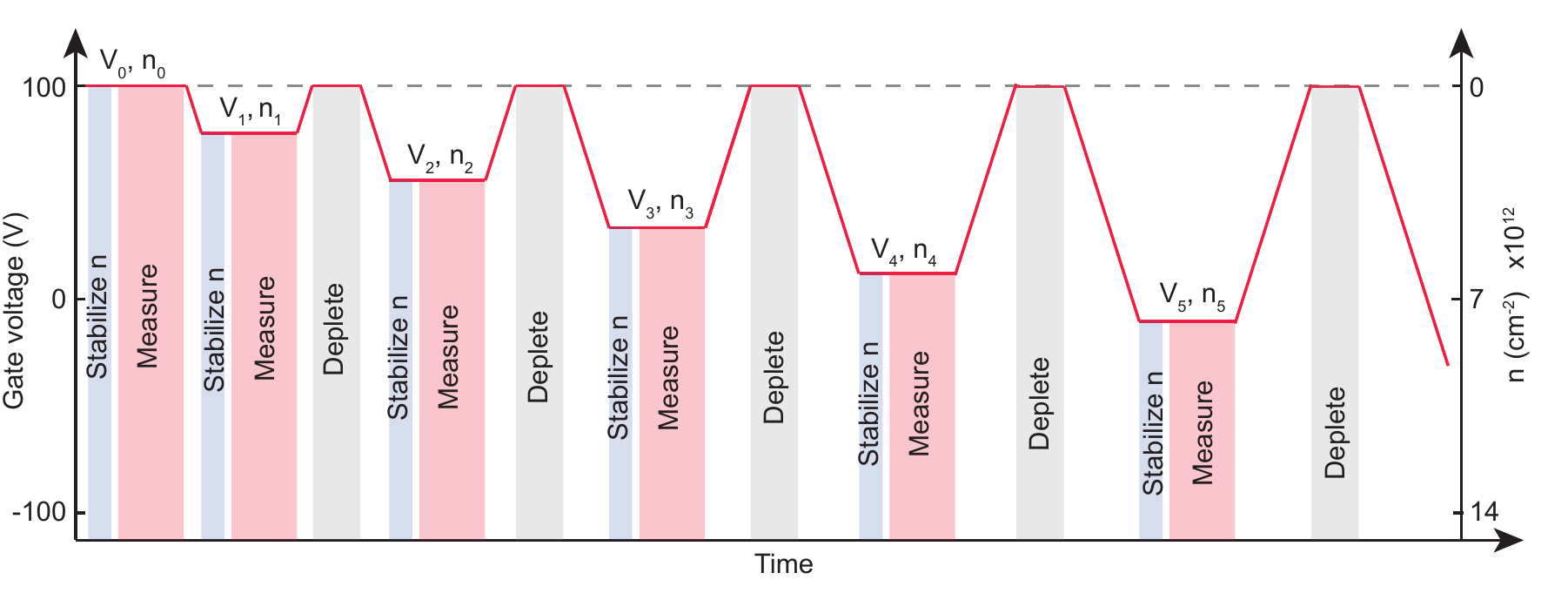}
\caption{The gate-sweep procedure. In order to circumvent issues stemming from the photo-doping effect, the sample is always depleted before electrons are injected. In this way, a given carrier density $n_j$ is reached at a specific gate voltage $V_j$. The stabilise, measure and deplete durations were 20~s, 1,200~s and 10~s, respectively.}
\label{gateSweep}
\end{figure*}

Upon laser illumination, charges become trapped at defects and at interfaces and screen the electric field across our capacitor~\cite{FabJu2014,Epping2017,Wang2010}. A direct consequence is that the electron density under the laser spot decreases with laser exposure time. This effect varies locally as it depends on the exact locations of the defects in the h-BN and on the location of imperfections within the van der Waals heterostructure. The carrier density in the MoS$_2$ monolayer therefore varies locally after a long laser exposure, even within the size of the focussed spot. In these experiments, very low laser powers are used. Nevertheless, even at these low powers, changes in the measured optical spectra occur on a slow time-scale of an hour on account of photo-doping. As the acquisition time needed to obtain a full dataset such as Fig.~2 of the main text is around ten hours, the problem of the photo-doping must be addressed.

In order to circumvent the problem of photo-doping, we sweep our gate voltage in a particular way, as depicted in Fig.~\ref{gateSweep}. The MoS$_2$ is always depleted after measuring at one density and before measuring at the next. This reinitialisation of the charge state of the monolayer is important as the measurement time at a specific electron density is then shorter than the time scale of the photo-doping effect.

\section{Additional PL data}
Here, additional PL data obtained on the same sample as in the main text are presented. We present colour-maps of the PL in the absence of magnetic field, $B_{z}=0.00$~T, and at $T=1.6$~K; and at a magnetic field of $B_z=9.00$~T and at $T=30$~K.
\subsection{Photoluminescence without an applied magnetic field}
\label{nomag}
\begin{figure*}[t]
\centering
\includegraphics[width=125mm]{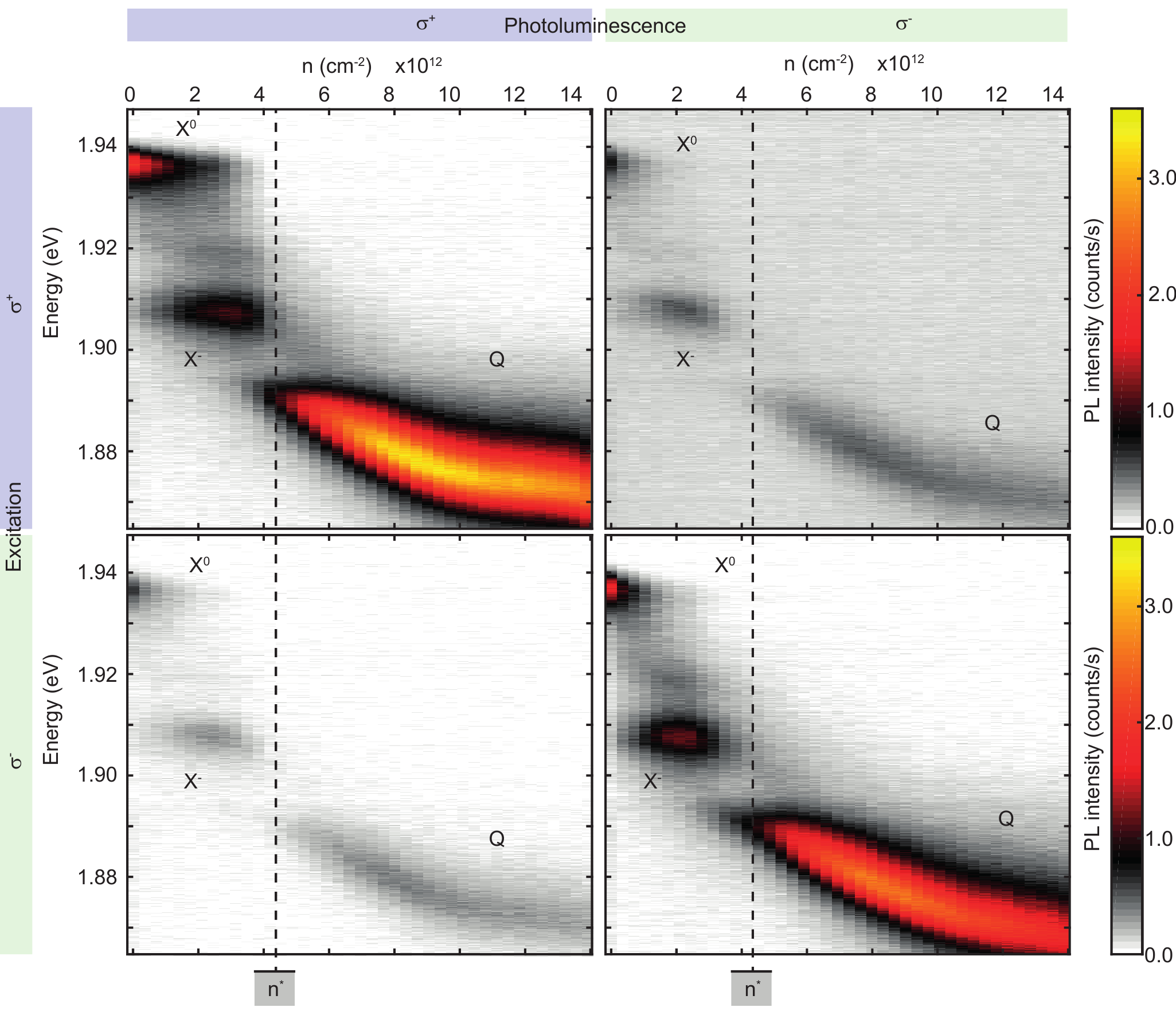}
\caption{Photoluminescence of the gated MoS$_2$ without magnetic field ($B_{z}=0.00$ T) at temperature $T=1.6$~K.}
\label{PL_0T}
\end{figure*}

Fig.~2 of the main text presents colour-maps of the PL as a function of the electron concentration in an external magnetic field of $B_{z}=9.00$~T and at a temperature of $T=1.6$~K. In Fig.~\ref{PL_0T}, we show PL from the same sample, but at $B_z=0.00$~T and $T=1.6$~K. No hints of a broken-symmetry phase can be detected in the absence of magnetic field, as is also the case in the absorption data~\cite{Roch2019}. However, even at $B_z=0.00$~T, the PL energy jumps abruptly by 15~meV at the electron density $n_{\rm c}=3.0 \times 10^{12}$~cm$^{-2}$, as at $B_z=9.00$~T. This similarity in the PL at $B_z=0.00$~T with respect to the PL at $B_z=9.00$~T can be understood in terms of fluctuations. In the absence of a magnetic field, the spin-up and the spin-down polarised states in the ferromagnetic phase are degenerate. Fluctuations of the sign of the magnetisation on length scales smaller than the focus of our laser spot and on time scales smaller than the integration time of the experiment prevent us from detecting one of these phases. Both spin-up and spin-down phases contribute equally to the PL signal. Nevertheless, there is an abrupt change in both $E_{\rm PL}$ and $E_{\rm PL}-E_{\rm A}$ at the critical density. This tells us that at $B_z=0.00$~T and $T=1.6$~K there is still a change in spin ordering at the critical density, $n_{\rm c}$. For $n<n_{\rm c}$, there are strong correlations between one spin and the next at each moment in time; for $n>n_{\rm c}$ these correlations disappear.
\subsection{Photoluminescence at elevated temperature}
\begin{figure*}[t]
\centering
\includegraphics[width=125mm]{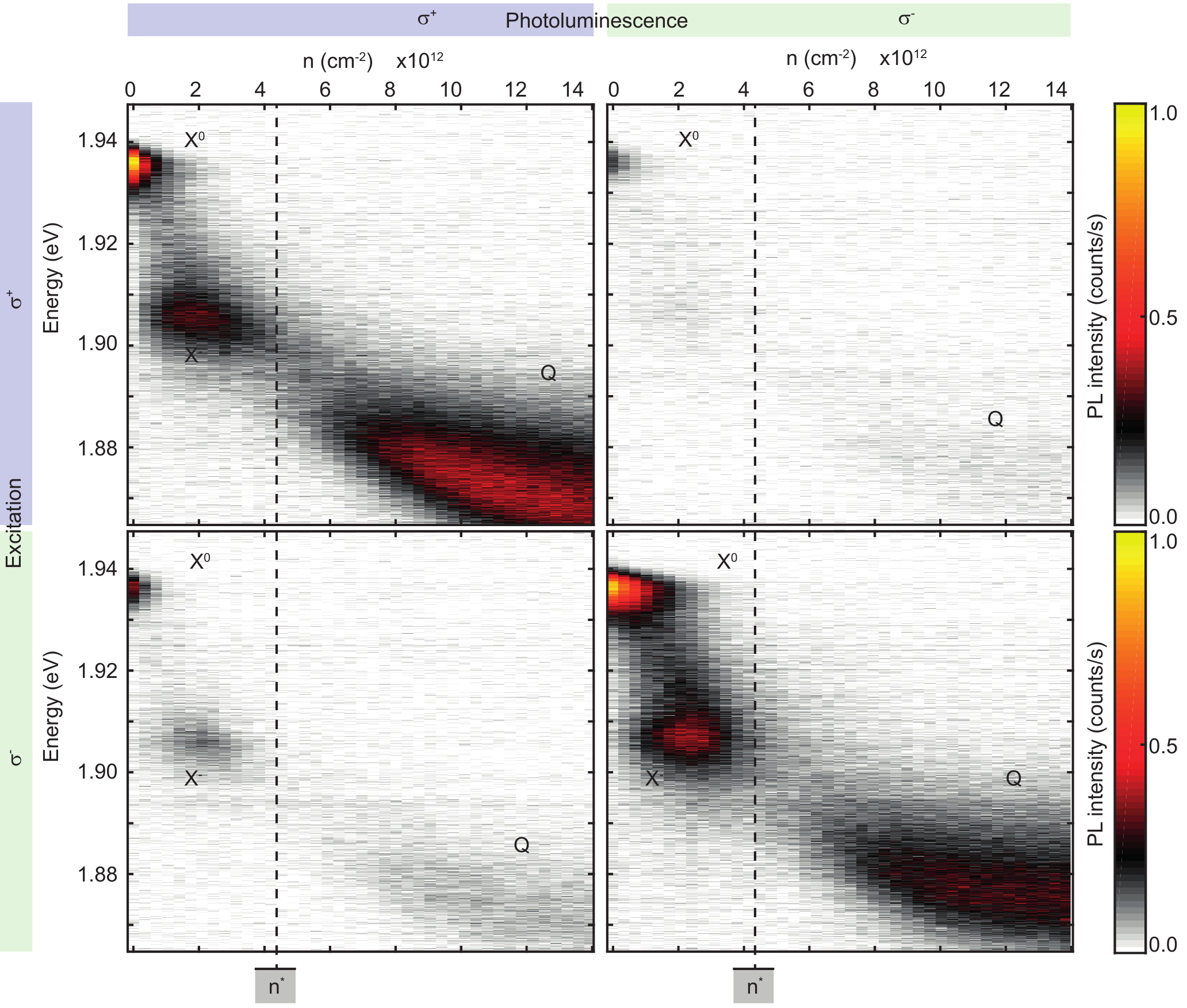}
\caption{Photoluminescence of the gated MoS$_2$ at magnetic field $B_z=9.00$~T and temperature $T=30$~K.}
\label{PL_T30K}
\end{figure*}

Fig.~\ref{PL_T30K} shows the PL data from the same sample as in the main text, at $B_{z}=9.00$~T but at an elevated temperature of $T=30$~K. The contrast between the two collection polarisations is less pronounced than that obtained at lower temperature ($T=1.6$~K, Fig.~2 of the main text). The decrease in polarisation contrast was also observed on absorption data in monolayer MoS$_2$~\cite{Roch2019}. However, at $T=30$~K as at  $T=1.6$~K, a pronounced jump of 15~meV is also observed in the PL on increasing the electron density (Fig.~\ref{PL_T30K}). In the light of fluctuations, as discussed in Section~\ref{nomag}, the decrease in circular dichroism without a deep change in the optical spectrum is understood in terms of fluctuations. Application of a magnetic field $B_{z}=9.00$~T at $T=1.6$~K stabilises the spin in the ferromagnetic phase: fluctuations from one spin-state to the other become rare. At $B_{z}=9.00$~T but at $T=30$~K, there are thermally driven flucations from one spin-state to the other. Nevertheless, the PL spectrum tells us that the phase transition still exists: there are strong local spin-correlations for $n<n_{\rm c}$; but no spin-correlations for $n>n_{\rm c}$. We anticipate that application of a larger magnetic field would stabilise the spin in the ferromagnetic phase for $n>n_{\rm c}$.

\begin{figure}[t]
\centering
\includegraphics[width=0.7\columnwidth]{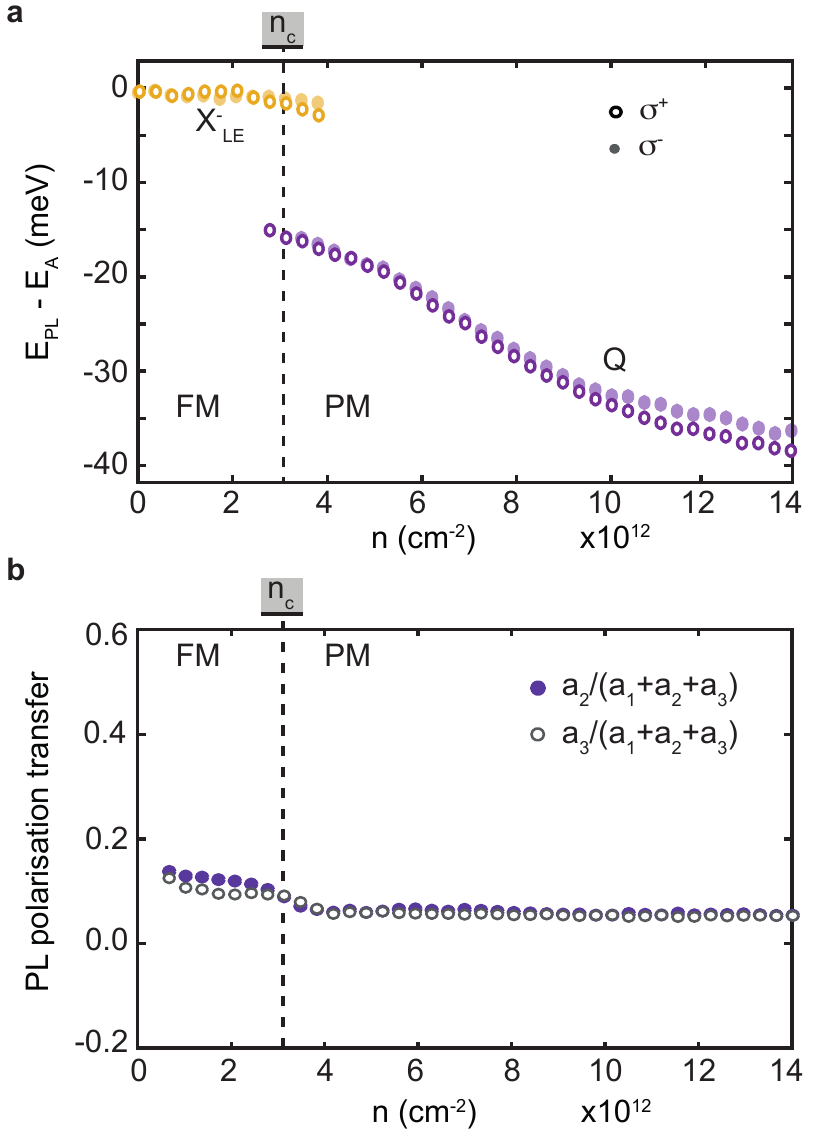}
\caption{(a) $E_{\rm PL}-E_{\rm A}$, (b) PL polarisation transfer, at $B_z=0.00$~T and $T=1.6$~K.}
\label{PLtransfer_B0}
\end{figure}

\begin{figure}[t]
\centering
\includegraphics[width=0.7\columnwidth]{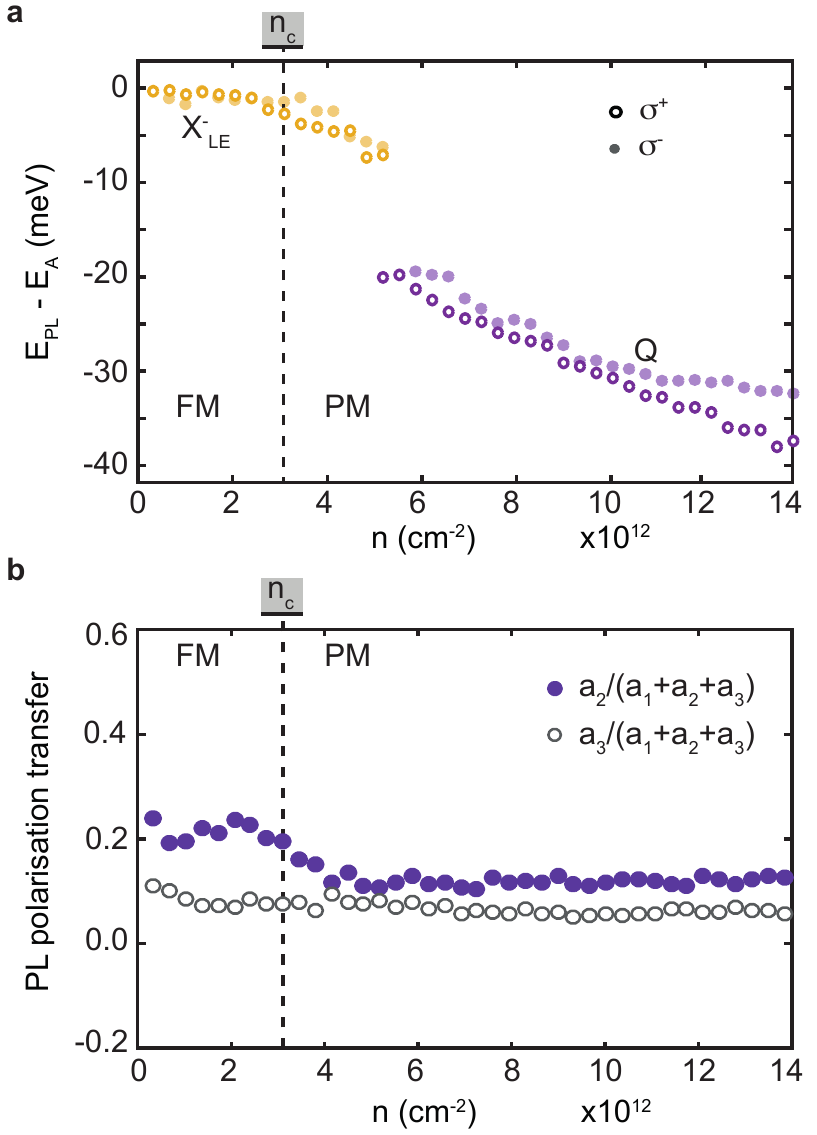}
\caption{(a) $E_{\rm PL}-E_{\rm A}$, (b) PL polarisation transfer, at $B_z=9.00$~T and $T=30$~K.}
\label{PLtransfer_T30}
\end{figure}

Both data-sets ($B_{z}=0.00$~T, $T=1.6$~K; $B_{z}=9.00$~T, $T=30$~K) can be analysed in the same way as the $B_{z}=9.00$~T, $T=1.6$~K data-set, i.e.\ in terms of $E_{\rm PL}-E_{\rm A}$ and the PL polarisation transfer. The results are shown in Fig.~\ref{PLtransfer_B0} and Fig.~\ref{PLtransfer_T30}, respectively.

\section{Absorption measurement}
The absorption measurements were carried out as in Roch {\em et al.}~\cite{Roch2019}. The differential reflectivity is recorded using extremely weak, broadband illumination. The optical susceptibility is determined from the differential reflectivity. The trion and Q-peak spectral positions were determined by fitting the spectral resonances to a Lorentzian function. This procedure determines the energy of the absorption resonance, $E_{\rm A}$.

\end{document}